\documentstyle[12pt]{article}
\newcommand{\HV}{\hat{V}}
\newcommand{\HH}{\hat{H}}
\newcommand{\HT}{\hat{T}_\varepsilon}
\newcommand{\HG}{\hat{G}_\varepsilon}
\newcommand{\HGO}{\hat{G}_\varepsilon^{(0)}}

\newcommand{\ov}{\overline}

\begin{document}%
%\baselineskip 22pt plus 2pt

%%%%%%%%%%%%%%%Title page%%%%%%%%%%%%%%%%
\title{\Large \bf
Analytical properties of scattering amplitudes\\
in one-dimensional quantum theory}
\author{
{\normalsize\sc M. S. Marinov$^{*,**}$ and Bilha Segev$^*$ }\\
{\normalsize\it
$^*$Department of Physics, Technion-Israel Institute of Technology}\\
{\normalsize\it Haifa 32000, Israel}\\
{\normalsize\it
$^{**}$Max-Planck-Institut f\"{u}r Physik and Astrophysik}\\
{\normalsize\it Werner-Heisenberg-Institut, M\"{u}nchen, Germany }\\
}

\date{        }

\maketitle
%%%%%%%%%%%%%%%%%%%%%%%%%%%%%%%%%%%%%%%%%%%%%%
{\begin{abstract}
One-dimensional quantum scattering from a local potential barrier
is considered.
Analytical properties of the scattering amplitudes
have been investigated by means of the integral equations
equivalent to the Schr\"{o}dinger equations. 
The transition and reflection amplitudes are expressed
in terms of two complex functions of the incident energy, 
which are similar to the Jost function 
in the partial-wave scattering. These functions are entire for finite-range
potentials and meromorphic for exponentially decreasing potentials. 
The analytical properties result from locality of the potential in the
wave equation and represent the effect of causality in time dependence
of the scattering process.
\end{abstract}

{\em PACS numbers}: 0230, 0380, 1155, 0365.

{\em Short title}: Analytical properties of scattering amplitudes 

\newpage
\section{Introduction}

The problem of the {\em tunneling time} has been attracting a
considerable attention for decades
\cite{Buttiker,Hauge&Stovneng,Landauer&Martin,Steinberg}.
It is indeed important to understand the effect of
{\em causality} upon the particle and wave propagation.
The problem is that, in the conventional time-independent
formalism, the causality manifests itself indirectly, namely, in
analytical properties of the transition amplitudes as functions of
(complex) energy. The relations between causality and analyticity
were under an intensive investigation in the 60-ties, when the concept
of the $S$-matrix dominated in particle physics\cite{eden,peres}.
At that time, however, the analysis was aimed mainly at
three-dimensional scattering processes, for central-symmetric potentials
in particular\cite{regge,newton-book}.
Partial-wave scattering amplitudes in the complex energy plane were 
considered also in the theory of multi-channel nuclear reactions\cite{mello}.
Analytical properties of the scattering matrix have been used also in
other fields. For example, in the theory of multi-terminal mesoscopic 
conductance the properties of the multi-probe $S$-matrix were 
used\cite{shepard} to obtain the conductance and to establish the
time ordering of the incoming and outgoing lead states. In another
work\cite{buttiker93}, the low-frequency behaviour of dynamic conductance
was related to the phase-delay times for the carrier transmission and 
reflection, which are given by the energy derivatives of the $S$-matrix 
elements. The most imporatnt feature of the transition amplitudes for 
various physical processes in the energy representation is their 
analyticity in the upper half of the complex plane. Sometimes one can 
find out more about singularities in the lower half-plane, investigating
dynamical equations specific for the physical problem, like the 
Schr\"{o}dinger equation for scattering. 

The one-dimensional `scattering', i.e. the potential-barrier problem,
is somewhat more complicated than the scattering off a force center,
since the system has two channels, corresponding to two waves running
in the opposite directions from the potential region in the final
state. Thus instead of one analytical function, the partial wave
scattering amplitude, one deals with two analytical functions,
transmission and reflection amplitudes. (The unitarity condition holds in
both cases.)
The familiar arguments of the scattering theory 
must be extended properly to the one-dimensional case.
The analytical properties of the one-dimensional
$S$ matrix were discussed, in particular, 
by Faddeev\cite{faddeev59,faddeev64} and Newton\cite{newton-pap} 
in view of the inverse problem.
The singularities in the complex energy plane, caused by bound and 
quasi-bound states were considered more recently
\cite{Gurvitz,Azbel,Bianchi}. 

The purpose of the present work is to investigate the location and 
character of singularities in the scattering amplitudes, owing
to the potential shape. The investigation is based upon the 
Schr\"{o}dinger equation with a local potential. That enables one to reveal
general features of the amplitudes, depending on the character
of vanishing of the potential outside its domain. 
The analytical properties are essential 
for the application to a space-time picture of the barrier transmission.

%%%%Contents
In Section 2, the $2\times 2$ scattering and transition matrices are
introduced and related to the resolvent of the Schr\"{o}dinger
operator. Next, in Section 3, two complex functions are defined
for the potential barrier problem, which are related to elements
of the monodromy matrix. Their role is similar to that of
the Jost function in the $S$-wave potential scattering.
These functions have nice analytical properties, which are
proven in Section 4 by means of the Volterra-type integral equations.
It is shown, in particular, that the functions have no singularities
in the whole complex energy plane (except for the infinity), if
the potential has a finite range. Singularities appear if the potential
behaves exponentially in the asymptotics, and the slope of the exponent
determines the distance to the singularies nearest to the real energy 
axis. Some examples are given in Appendix.

\section{Transition operator and the $S$-matrix}
The evolution operator is given by the Laplace transform
of the resolvent of the Hamiltonian $\HH$,
   \begin{equation} %(1)
\exp(-it\HH)=\frac{1}{2\pi i}\int_{\Gamma_\infty}
\HG e^{-i\varepsilon t}d\varepsilon,\;\;\;
\HG\equiv (\HH-\varepsilon)^{-1}.
      \end{equation}
Here $\Gamma_\infty$ is the contour in the complex $\varepsilon$-plane,
running from $-\infty$ to  $+\infty$ {\em above} the real axis, where the
singularities are situated.
In scattering problems, the Hamiltonian is a sum of
a potential operator and the kinetic energy, having the eigen-vectors
describing free particle states,
  \begin{equation} %(2)
\HH \equiv\HH_0+\HV, \;\;\;
\HH_0|{\bf k}>=\epsilon (k)|{\bf k}>.
      \end{equation}
In the nonrelativistic scattering theory, we shall use the units where 
 $\epsilon (k)=k^2$.
The transition operator $\HT$ is introduced as follows,
  \begin{equation} %(3)
\HG=\HGO-\HGO\HT\HGO,\;\;\;\HGO\equiv (\HH_0-\varepsilon)^{-1}.
      \end{equation}
It can be expressed directly in terms of the resolvent $\HG$,
     \begin{equation}%(4)
\HT=\HV-\HV\HG\HV,
      \end{equation}
As follows from the time-inversion symmetry (the reciprocity principle),
      \begin{equation}%(5)
<{\bf k}|\HT|{\bf k}_0>=<-{\bf k}_0|\HT|-{\bf k}>.
    \end{equation}
It is easy to see, using
the standard definition of the scattering operator $\hat{S}$,
that its matrix elements are expressed
in terms of the transition operator on the energy shell,
  \begin{eqnarray} %(6)
<{\bf k}|\hat{S}|{\bf k}_0>\equiv\lim_{t\rightarrow\infty}
<{\bf k}|e^{\frac{i}{2}t\HH_0}e^{-it\HH}e^{\frac{i}{2}t\HH_0}|{\bf k}_0>
\nonumber\\
=\delta({\bf k-k}_0)-2\pi i\delta[\epsilon (k)-\epsilon (k_0)]
T_{\nu\nu_0}(\varepsilon).
         \end{eqnarray}
Here the scattering amplitude is given by
  \begin{equation} %(7)
T_{\nu\nu_0}(\varepsilon) \equiv<{\bf k}|\HT|{\bf k}_0>,\;\;\;
\epsilon (k)=\varepsilon=\epsilon (k_0),
      \end{equation}
where $\nu\equiv {\bf k}/k$, and the standard normalization is used:
$<{\bf k}|{\bf k}_0>=\delta({\bf k-k}_0)$.

In the one-dimensional case  ${\bf k}=\nu k$, where $\nu=\pm 1$,
corresponding to two possible
directions of motion for a given energy, and
     \begin{equation}%(8)
\delta({\bf k-k}_0)=v\delta[\epsilon(k)-\epsilon(k_0)]\delta_{\nu\nu_0},\;\;\;
v=d\epsilon/dk.
     \end{equation}
The elements of $\hat{S}$ and $\hat{T}$ (on the energy shell)
are given by $2\times 2$ matrices $S$ and $T$,
     \begin{equation} %(9)
<{\bf k}|\hat{S}|{\bf k}_0>=v\delta[\epsilon(k)-\epsilon(k_0)]S_{\nu\nu_0},
\;\;\;S\equiv I-\frac{2\pi i}{v}T,
     \end{equation}
 By definition, if $\hat{S}$ exists, it is a unitary operator.
This fact implies a unitarity condition
on the scattering amplitude, which reads
     \begin{equation}%(10)
SS^\dagger=I,\;\;\;
\frac{1}{2\pi i}(T-T^\dagger)=-\frac{1}{v}TT^\dagger.
      \end{equation}
As we will show,   
the analytical properties of $T(\varepsilon)$ follow from locality
of the potential, and Eq. (4).

\section{General properties of the transition amplitudes}

We consider the Schr\"{o}dinger equation,
     \begin{equation}%(11)
\HH\psi=\kappa^2\psi,\;\;\;\;\HH=-(d/dx)^2+V(x),
     \end{equation}
where $-\infty<x<\infty$, and the potential is local, i.e.
$|V(x)|=o(1/x)$ as $|x|\rightarrow\infty$. In the coordinate
representation, the resolvent can be expressed
in terms of two fundamental solutions of the Schr\"{o}dinger equation,
$y_\pm(x)$, satisfying the proper boundary conditions at $\pm\infty$,
respectively, 
\begin{eqnarray}%(12-13)
<x|\HG|x_0>=\frac{y_-(x_<)y_+(x_>)}{w(y_-,y_+)},
\;\;\;x_{>/<}={\rm max/min}(x,x_0),\\
y_\pm(x)\rightarrow e^{\pm i\kappa x},\;\;\;{\rm as}\;x\rightarrow
\pm\infty,\\w(y_-,y_+)\equiv y_-'y_+-y_-y_+'={\rm const}.\nonumber
     \end{eqnarray}
As soon as $\varepsilon\equiv\kappa^2$ is introduced in the 
Laplace transform
(1) for the {\em upper} half of the complex plane, the solutions $y_\pm$
defined above vanish at $\pm\infty$, respectively. Thus one has the
properly defined resolvent for the elliptic operator $\HH$ satisfying
the Sommerfeld radiation condition at infinity\cite{sommer}.

For large $|x|$, where the potential vanishes, the asymptotics of the
fundamental solutions are given by
       \begin{eqnarray}%(14)
y_-(x)=
a e^{-i\kappa x}+be^{i\kappa x}\;\;\;\;{\rm for}\;x\rightarrow+\infty\\
y_+(x)=b'e^{-i\kappa x}+ce^{i\kappa x}\;\;\;\;{\rm for}\;x\rightarrow
-\infty.\nonumber
       \end{eqnarray}
In principle, the solutions are defined by (13) for Re$\kappa>0$ and
Im$\kappa\rightarrow +0$, yet the analytical continuation to the whole
complex $\kappa$-plane is considered in the following.
For real potentials $V(x)$, the complex conjugate functions
$\overline{y_\pm^\kappa(x)}$
are also solutions of the Schr\"{o}dinger equation,
satisfying the boundary conditions conjugate to (13).
Calculating the Wronskians (which are independent of $x$) at
$x\rightarrow\pm\infty$ for various pairs of the solutions,
one gets a number of relations between the complex parameters $a,b,b',c$:
   \begin{eqnarray}%(15-17)
w(y_-,y_+)=-2i\kappa a=-2i\kappa c\;\leadsto a=c,\\
w(y_-,\bar{y}_-)={\rm const}\leadsto |a|^2-|b|^2=1,\\
w(y_+,\bar{y}_+)={\rm const}\leadsto |c|^2-|b'|^2=1,\nonumber\\
w(y_-,\bar{y}_+)={\rm const}\leadsto b'=-\bar{b}.
    \end{eqnarray}
Expressing $\overline{y_\pm}$ in terms of the fundamental solutions, one has
       \begin{eqnarray}%(18)
\bar{y}_-=\frac{\bar{b}}{a}y_-+\frac{1}{a}y_+,\;\;\;
\bar{y}_+=\frac{1}{a}y_--\frac{b}{a}y_+.
       \end{eqnarray}
The analytical continuation of these solutions to the complex $\kappa$ plane, 
by $\overline{y_\pm^\kappa(x)}\equiv y_\pm^{-\bar{\kappa}}(x)$, 
implies a symmetry of $a(\kappa)$ and $b(\kappa)$ with respect to the  
imaginary $\kappa$-axis,
   \begin{equation}%(19)
\ov{a(\kappa)}=a(-\bar{\kappa}), \;\;\;\; \ov{b(\kappa)}=b(-\bar{\kappa}).
    \end{equation}

Thus, the asymptotics of the solutions depend only on two complex functions 
$a(\kappa)$ and $b(\kappa)$, satisfying one real condition (16), 
and subject to the symmetry (19). The transition amplitudes, elements of 
the $S$ matrix and $T$ matrix, and of the monodromy matrix \cite{arnold},
are given in terms of these two functions.

If the potential is displaced, $b$ gets a phase shift,
    \begin{equation}%(20)
V(x)\rightarrow V(x-d)\;
\leadsto a\rightarrow a,\;b\rightarrow be^{-2i\kappa d}.
    \end{equation}
For symmetric potentials one gets an additional relation,
    \begin{equation}%(21)
V(x)\equiv V(-x)\;\leadsto y_-(-x)\equiv y_+(x):\;\;\; a=c,\;b'=b,
    \end{equation}
so that $b$ is pure imaginary, in view of (17).

As soon as the resolvent is known from (12),
the elements of the transition
operator in the momentum representation are obtained immediately, by (4),
      \begin{equation}%(22)
<k|\HT|k_0>=\tilde{V}(q)-\frac{1}{2\pi}\int\!\!\int dxdx_0
e^{-ikx+ik_0x_0}V(x)<x|\HG|x_0>V(x_0),
    \end{equation}
where $q=k-k_0$, and
     \begin{equation}%(23)
\tilde{V}(q)=\frac{1}{2\pi}\int^{+\infty}_{-\infty}V(x)e^{-iqx}dx.
     \end{equation}
The double Fourier transform in (22) is performed by means of the
Schr\"{o}dinger equation, $Vy=y''+\kappa^2y$, leading
to the following integrals
      \begin{eqnarray}%(24-25)
\int_{-\infty}^xe^{-ik\xi}V(\xi)y_-(\xi)d\xi
=(\eta'_-+ik\eta_-)e^{-ikx}\nonumber\\
+(\kappa^2-k^2)\int_{-\infty}^xe^{-ik\xi}\eta_-(\xi)d\xi,\\
\int^{\infty}_xe^{-ik\xi}V(\xi)y_+(\xi)d\xi=
-(\eta'_++ik\eta_+)e^{-ikx}\nonumber\\
+(\kappa^2-k^2)\int^{\infty}_xe^{-ik\xi}\eta_+(\xi)d\xi,
    \end{eqnarray}
where we have introduced the 
functions $\eta_\pm(x)$ vanishing at $\pm\infty$,
      \begin{equation}%(26)
\eta_\pm(x)\equiv y_\pm(x)-e^{\pm i\kappa x}.
     \end{equation}
The result is
    \begin{eqnarray}%(27)
<k|\HT|k_0>=\frac{1}{2\pi iw}\int^{\infty}_{-\infty}dxe^{-iqx}V(x)
\left[ (\kappa-\frac{q}{2})y_+(x)e^{-i\kappa x}+
 (\kappa+\frac{q}{2})y_-(x)e^{i\kappa x}\right]\nonumber\\
-\frac{\kappa^2-\frac{1}{2}(k^2+k_0^2)}{2\pi w}
\int^{\infty}_{-\infty}dxe^{-iqx}(\eta_+\eta'_--\eta_-\eta'_+)\nonumber\\
-(\kappa^2-k^2)(\kappa^2-k_0^2)
\frac{1}{2\pi}\int\!\!\int dxdx_0e^{-ikx+ik_0x_0}g(x,x_0),
     \end{eqnarray}
where $w=-2i\kappa a$, and
      \begin{equation}%(28)
g(x,x_0)\equiv \eta_-(x_<)\eta_+(x_>)/w.
    \end{equation}

On the energy shell where $k^2=\kappa^2=k_0^2$, the contributions
from the integrals vanish in Eqs. (24-27), and one has
      \begin{equation}%(29)
T=\frac{1}{2\pi a}
\left(\begin{array}{cc}
\alpha & \beta\\
\bar{\beta} & \alpha
\end{array}\right),\;\;\;
S=\frac{1}{a}
\left(\begin{array}{cc}
1 & b \\
-\bar{b} & 1
\end{array}\right),
    \end{equation}
where
      \begin{eqnarray}%(30-31)   
\alpha\equiv \int^{\infty}_{-\infty}dxe^{-i\kappa x}  V(x)y_+(x)=
 \int^{\infty}_{-\infty}dxe^{i\kappa x}  V(x)y_-(x),\\
\beta\equiv \int^{\infty}_{-\infty}dxe^{i\kappa x}  V(x)y_+(x)= 
\int^{\infty}_{-\infty}dxe^{-i\kappa x}  V(x)y_-(x),\nonumber\\
a\equiv 1-\frac{\alpha}{2i\kappa},\;\;\;b\equiv\frac{\beta}{2i\kappa}.
     \end{eqnarray}
The functions $\alpha(\kappa)$ and $\beta(\kappa)$ are free of
a pole at $\kappa=0$, and are related by the unitarity condition,
      \begin{equation}%(32)
\alpha-\bar{\alpha}=
\frac{i}{2\kappa}(\alpha\bar{\alpha}-\beta\bar{\beta}).
    \end{equation}
The transmission and reflection amplitudes are
$S_{++}=1/a$ and $S_{+-}=b/a$, respectively, and $\det S=\bar{a}/a$.
(The latter equality supports the analogy of $a$ to the Jost function.
If there is no reflection, $b=0$, then $a=e^{-i\delta}$, where
 $\delta(\kappa)$ is a real phase shift.) Besides, if the potential is even, 
the matrices $S$ and $T$ are symmetrical.

Note that because of the relations (15)-(17) the monodromy matrix 
\cite{arnold} composed of $a,b$ is quasi-unitary,
     \begin{equation}%(33)
 M=\left(\begin{array}{cc}
 \bar{a} & -\bar{b} \\
 -b & a
 \end{array}\right),\;\;\;MEM^\dagger=E,\;\;\;
 E=\left(\begin{array}{cc}
 1 & 0 \\
 0 & -1
 \end{array}\right).
     \end{equation}
Under a displacement of the potential, Eq. (20), $M$ is transformed to
      \begin{equation}%(34)
M\rightarrow DMD^\dagger,\;\;\;
D=\left(\begin{array}{cc}
1 & 0 \\
0 & e^{-2i\kappa d}
\end{array}\right).
    \end{equation}
If $V(x)=V_1(x-d_1)+V_2(x-d_2)$, $d_2>d_1$, 
and the potential domain consists
of two intervals separated by a forceless gap, one has the following
superposition rule,
      \begin{eqnarray}%(35)
M=D_1M_1D_1^\dagger D_2M_2D_2^\dagger:\nonumber\\
a=a_1a_2+b_1\bar{b}_2e^{2i\kappa(d_2-d_1)},\;
b=a_1b_2e^{-2i\kappa d_2}+\bar{a}_2b_1e^{-2i\kappa d_1},
    \end{eqnarray}
which is a consequence of Eqs. (14) and (20).

\section{Integral equations and the analytical properties}
\subsection{Volterra equation and series solutions}
Analytical properties of the transition amplitudes can be derived
from the integral equation satisfied by the fundamental solutions
$y_\pm(x)$,
    \begin{equation}%(36)
y_\pm(x) =e^{\pm i\kappa x}+\frac{1}{\kappa}
\int_{\pm\infty}^x\sin\kappa(x-\xi)V(\xi)y_\pm(\xi)d\xi.
    \end{equation}
These equations are of the Volterra type, so the solution exists and
admits an analytical continuation to complex $\kappa$ for local
potentials. In order to separate asymptotic  oscillations of the solutions,
let us introduce new functions (the integrals are evaluated by Eqs. (24-25)
for $k=\pm\kappa$),
      \begin{eqnarray}%(37-38)
A_-(x)\equiv\int_{-\infty}^xe^{i\kappa \xi}V(\xi)y_-(\xi)d\xi
=(y'_--i\kappa y_-)e^{i\kappa x}+2i\kappa,\nonumber\\
B_-(x)\equiv\int_{-\infty}^xe^{-i\kappa \xi}V(\xi)y_-(\xi)d\xi
=(y'_-+i\kappa y_-)e^{-i\kappa x},\\
A_+(x)\equiv\int^{\infty}_xe^{-i\kappa \xi}V(\xi)y_+(\xi)d\xi
=-(y'_++i\kappa y_+)e^{-i\kappa x}+2i\kappa,\nonumber\\
B_+(x)\equiv\int^{\infty}_xe^{i\kappa \xi}V(\xi)y_+(\xi)d\xi
=-(y'_+-i\kappa y_+)e^{i\kappa x}.
    \end{eqnarray}
It is easy to see that
     \begin{eqnarray}%(39)
\frac{dy_\pm}{dx}= \pm i\kappa\left(
  1-\frac{1}{2i\kappa }A_\pm(x)\right)e^{\pm i\kappa x}
 \mp\frac{1}{2}B_\pm(x)e^{\mp i\kappa x},\nonumber\\
  y_\pm(x)=\left(1-\frac{1}{2i\kappa }A_\pm(x)\right)e^{\pm i\kappa x}
 +\frac{1}{2i\kappa }B_\pm(x)e^{\mp i\kappa x},
     \end{eqnarray}
so $A$ and $B$ have the definite limits
at $x\rightarrow\pm\infty$, cf. Eqs. (30),
      \begin{eqnarray}%(40)
\alpha\equiv A_+(-\infty)=A_-(+\infty),\;\;\;
\beta\equiv B_-(+\infty)=\overline{B_+(-\infty)},\\
A_\pm(\pm\infty)=0=B_\pm(\pm\infty).\nonumber
    \end{eqnarray}
The pairs of functions $(A,B)$ satisfy a system of first-order
differential equations with zero initial conditions at infinity.
Setting the equations into the integral form, one gets from (37),
in particular, for $A_-$ and $B_-$,
       \begin{eqnarray}%(41-42)
 A_-(x)=\int_{-\infty}^xV(\xi)\left(
 1-\frac{1}{2i\kappa }A_-(\xi)
 +\frac{1}{2i\kappa }B_-(\xi)e^{2i\kappa\xi}\right)d\xi,\\
 B_-(x)=\int_{-\infty}^xV(\xi)\left(
 (1-\frac{1}{2i\kappa}A_-(\xi))e^{-2i\kappa\xi}
 +\frac{1}{2i\kappa }B_-(\xi)\right)d\xi.
     \end{eqnarray}
This form is especially suitable for the perturbative expansion of
$\alpha$ and $\beta$, namely,
      \begin{eqnarray}%(43-44)
&&\alpha(\kappa)=\int^{+\infty}_{-\infty}\!\!V(\xi)d\xi\nonumber\\
&&+\frac{1}{\kappa}\int^{+\infty}_{-\infty}\!\!d\xi_2
\int^{\xi_2}_{-\infty}\!\!d\xi_1V(\xi_2)V(\xi_1)
e^{i\kappa(\xi_2-\xi_1)}\sin\kappa(\xi_2-\xi_1)+\cdots ,\\
&&\beta(\kappa)=\int^{+\infty}_{-\infty}\!\!V(\xi)e^{-2i\kappa\xi}d\xi
\nonumber\\
&&+\frac{1}{\kappa}\int^{+\infty}_{-\infty}\!\!
d\xi_2\int^{\xi_2}_{-\infty}\!\! d\xi_1V(\xi_2)V(\xi_1)
e^{-i\kappa(\xi_2+\xi_1)}\sin\kappa(\xi_2-\xi_1)+\cdots .
    \end{eqnarray}
Using these expansions, one gets a sort of the Pad\'{e} approximation
for the $S$-matrix in (29). Note that as $\kappa\rightarrow 0$, one has
$\alpha-\beta\rightarrow 0$ while $\alpha$ and $\beta$ are regular,
so expanding in powers of $\kappa$ one can get, for short-range
potentials, an analogue of the effective-range
approximation\cite{newton-book}.

\subsection{Analytical properties}
The perturbative series are also used to prove the analytical properties.
First of all, one can extend to Eq. (36) the standard arguments
of the scattering theory\cite{newton-book}, which are based upon the
inequality
    \begin{equation}%(45)
|\sin\kappa(x-\xi)|\leq C\frac{|\kappa x|}{1+|\kappa x|}
e^{|{\rm Im}\kappa|(x-\xi)},
    \end{equation}
where $x>\xi$, and $C$ is a constant.
Thus one proves that $y_{\pm}(\kappa)$ are analytical
in the domain in the complex $\kappa$-plane where 
      \begin{eqnarray}%(nonumber)
\int^{x}_{-\infty}
\exp\left[-\xi (|{\rm Im} \kappa| \pm {\rm Im}\kappa)\right]
V(\xi)d\xi < \infty. \nonumber
    \end{eqnarray}
In particular, if $V(x)\equiv 0$ for $x<x_-$, for some $x_-$, there is
no irregularity as $x\rightarrow -\infty$. 
Similarly, the limit $x\rightarrow +\infty$ is considered.
The fundamental solutions are analytical in $\kappa$,
as soon as these two limits are regular.

One may rather modify the method and apply it directly to the
functions we are interested in, given by Eqs. (41-42).
The substitution
   \begin{equation}%(46)
A_-(x)=f(x)e^{iw(x)},\;\;\;
B_-(x)=g(x)e^{-iw(x)},\;\;\;
w(x)\equiv\frac{1}{2\kappa}\int^x_{-\infty}d\xi V(\xi),
\end{equation}
eliminates the diagonal terms in the differential equations, and they
are reduced to 
   \begin{eqnarray}%(47-48)
f'=f_0+P_+(x)g,&\;\;\;f_0=V(x)e^{-iw(x)},\\
g'=g_0+P_-(x)f,&\;\;\;g_0=V(x)\exp[-2i\kappa x+iw(x)].
       \end{eqnarray}
Here
   \begin{equation}%(49)
P_\pm(x)\equiv\pm\frac{V(x)}{2i\kappa}e^{\pm 2i[\kappa x-w(x)]}.
   \end{equation}
and the initial conditions are $f(-\infty)=0=g(-\infty)$.
Note that $f(x)$ and $g(x)$ have limits as $x\rightarrow\infty$,
which are $\alpha$ and $\beta$, up to conjugate phase shifts, provided that
the potential is integrable, and $w(\infty)$ is finite.
The solution to Eqs. (47-48) is given by the series,
   \begin{eqnarray}%(50-51)
f=\sum_{n=1}^\infty f_n,\;\;\;
f_1(x)=\int_{-\infty}^x d\xi f_0(\xi),\;\;\;
f_{n+1}(x)=\int_{-\infty}^x d\xi P_+(\xi)g_n(\xi),\\
g=\sum_{n=1}^\infty g_n,\;\;\;
g_1(x)=\int_{-\infty}^x d\xi g_0(\xi),\;\;\;
g_{n+1}(x)=\int_{-\infty}^x d\xi P_-(\xi)f_n(\xi).
       \end{eqnarray}

Upper bounds for $f_n$ and $g_n$ for positive 
and integrable potentials, can be obtained  by iteration.
From Eqs. (50-51), one gets for even $n$,
   \begin{eqnarray} %52 
f_n(x)=\int_{\Delta_n}P_+(\xi_{n-1}) P_-(\xi_{n-2}) ... P_+(\xi_1) g_0(\xi_0)
 \prod^{n-1}_{m=0}d\xi_m  , \nonumber \\
g_n(x)=\int_{\Delta_n}P_-(\xi_{n-1}) P_+(\xi_{n-2}) ... P_-(\xi_1) f_0(\xi_0)
 \prod^{n-1}_{m=0}d\xi_m,
     \end{eqnarray}
and for odd $n$,
   \begin{eqnarray} %53
f_n(x)=\int_{\Delta_n}P_+(\xi_{n-1}) P_-(\xi_{n-2}) ... P_-(\xi_1) f_0(\xi_0)
\prod^{n-1}_{m=0}d\xi_m, \nonumber \\ 
g_n(x)=\int_{\Delta_n}P_-(\xi_{n-1}) P_+(\xi_{n-2}) ... P_+(\xi_1) g_0(\xi_0) 
\prod^{n-1}_{m=0}d\xi_m.
      \end{eqnarray}
The integrations take place in the domain 
$\Delta_n:\;-\infty<\xi_0<\xi_1<\cdots <\xi_{n-1}<x$.
If we assume that $V(x)\geq 0$, $\kappa w(x)$ is a real, bounded and 
non-decreasing function of $x$.
For Im $\kappa > 0$, we shall use the following inequalities,
   \begin{eqnarray}%(54)
|P_+(\xi_i)P_-(\xi_j)| &\leq& 
\frac{V(\xi_i)}{|2\kappa|}\frac{V(\xi_j)}{|2\kappa|},\;\;\; \xi_i> \xi_j,
 \nonumber \\
|P_+(\xi_1)g_0(\xi_0)| &\leq& 
\frac{V(\xi_1)}{|2\kappa|}|f_0(\xi_0)|,  \nonumber \\
|f_0(\xi_0)|&\leq&V(\xi_0),  \nonumber \\
|P_-(\xi)|&\leq&\frac{V(\xi)}{|2\kappa|}|e^{- 2i\kappa \xi}|
|e^{ 2iw(x)}|, \;\;\; \xi<x.
       \end{eqnarray}
Similarly, for Im $\kappa < 0$,
   \begin{eqnarray}%(55)
|P_-(\xi_i)P_+(\xi_j)| &\leq& 
\frac{V(\xi_i)}{|2\kappa|}\frac{V(\xi_j)}{|2\kappa|},\;\;\; \xi_i> \xi_j,
 \nonumber \\
|P_-(\xi_1)f_0(\xi_0)| &\leq& 
\frac{V(\xi_1)}{|2\kappa|}|g_0(\xi_0)|,  \nonumber \\
|g_0(\xi_0)|&\leq&V(\xi_0)|e^{- 2i\kappa \xi_0}|,  \nonumber \\
|P_+(\xi)|&\leq&\frac{V(\xi)}{|2\kappa|}|e^{ 2i\kappa \xi}|
|e^{-2iw(x)}|, \;\;\; \xi<x.
       \end{eqnarray}
Using these inequalities one can show that, 
in the upper half of the complex $\kappa$-plane, 
   \begin{eqnarray}%(56)
|f_n(x)| &\leq& |2\kappa|\frac{|w(x)|^n}{n!},  \nonumber \\
|g_n(x) e^{-2iw(x)}| &\leq&|2\kappa|
\frac{|w(x)|^{n-1}}{(n-1)!}|u_- (x)|,
       \end{eqnarray}
while in the lower half of the complex 
$\kappa$-plane,
   \begin{eqnarray}%(57)
|f_n(x) e^{2iw(x)}| &\leq& |2\kappa|\frac{|w(x)|^{n-2}}{(n-2)!}
|u_- (x) u_+ (x)|
 \nonumber \\
|g_n(x)| &\leq&|2\kappa|
\frac{|w(x)|^{n-1}}{(n-1)!}|u_- (x)|,
       \end{eqnarray}
where
   \begin{equation}%(58)
u_\pm(x)\equiv
\frac{1}{2\kappa}\int^x_{-\infty}d\xi V(\xi)\exp(\pm 2i\kappa \xi).
    \end{equation}
It is assumed that the integrals exist for real $\kappa$ and have definite
limits as $x\rightarrow\infty$ (the Fourier transform of $V$).

For positive and integrable potentials, $f(x)$ and $g(x)$, 
and thus $\alpha(\kappa)$ and $\beta(\kappa)$, are given in terms of
infinite series which are absolutely convergent in the domain where
the corresponding Fourier transforms of $V(x)$ exist, Eq. (58).
Thus $\alpha(\kappa)$ and $a(\kappa)$  
are analytical in the upper half $\kappa$-plane.
The singularities of  $\alpha(\kappa)$ and $\beta(\kappa)$
appear if the integrals in Eq. (58) diverge as $x\rightarrow\infty$,
which may happen only at finite distances below the real axis.

\subsection{Finite-range potentials}
The functions $\alpha(\kappa)$ and $\beta(\kappa)$ are entire, if
$V(x)=0$ outside an interval $(x_-,x_+)$.  
The analyticity is an immediate result of Eqs. (56-58), 
as $u_\pm$ are finite for potentials with a finite support.

A more direct proof, as well as additional information on 
the analytic continuation into the complex $\kappa$-plane, can be obtained
from explicit expressions for $a(\kappa)$ and $b(\kappa)$.
Let us introduce two real (for real $\kappa$)
solutions of the Schr\"{o}dinger equation, $z_0(x)$ and
$z_1(x)$, specified by the following initial conditions,
\begin{eqnarray}%(59)
z_0(x_-)=1,\;\;\;z_0'(x_-)=0,\nonumber\\
z_1(x_-)=0,\;\;\;z_1'(x_-)=1.
\end{eqnarray}
From the continuity of the wave function and of its first derivative 
at $x=x_\pm$, one gets the following 
expressions for $a$ and $b$,
\begin{eqnarray}%(60)
a=\frac{e^{i\kappa(x_+-x_-)}}{2i\kappa}
[-\zeta_0'+i\kappa(\zeta_0+\zeta_1')+\kappa^2\zeta_1],\nonumber\\
b=\frac{e^{i\kappa(x_++x_-)}}{2i\kappa}
[\zeta_0'-i\kappa(\zeta_0-\zeta_1')+\kappa^2\zeta_1],
\end{eqnarray}
where $\zeta_{0,1}\equiv z_{0,1}(x_+)$. As soon as $\zeta$ and $\zeta'$
are analytical functions of $\kappa^2$, by the Poincar\'{e} 
theorem\cite{regge}, $\alpha(\kappa)$ and $\beta(\kappa)$ 
are also analytical in the whole complex $\kappa$-plane.

For large $|\kappa|$ and smooth $V(x)$, one can use the semi-classical 
approximation, 
\begin{eqnarray}%(61)
\zeta_0=\sqrt{\frac{p_-}{p_+}}\cos\theta,\;\;\;
\zeta'_0=-\sqrt{p_-p_+}\sin\theta,\nonumber\\
\zeta_1=\frac{1}{\sqrt{p_-p_+}}\sin\theta,\;\;\;
\zeta'_1=\sqrt{\frac{p_+}{p_-}}\cos\theta,
\end{eqnarray}
where
\begin{equation}%(62)
\theta(\kappa)\equiv\int^{x_+}_{x_-}\sqrt{\kappa^2-V(x)}dx,\;\;\;
p_\pm\equiv\sqrt{\kappa^2-V(x_\pm)}.
\end{equation}
In this approximation one gets
\begin{eqnarray}%(63)
a=\frac{e^{i\kappa(x_+-x_-)}}{2i\kappa\sqrt{p_-p_+}}
[(\kappa^2+p_-p_+)\sin\theta+i\kappa(p_-+p_+)\cos\theta],\nonumber\\
b=\frac{e^{i\kappa(x_++x_-)}}{2i\kappa\sqrt{p_-p_+}}
[(\kappa^2-p_-p_+)\sin\theta-i\kappa(p_--p_+)\cos\theta].
\end{eqnarray}
Note that this result is exact for the square-well barrier, Eq. (68),
and the analytical continuation to the complex plane is possible.
Asymptotical locations of zeroes of $a$ in the complex 
$\kappa$-plane are given by the equation
\begin{equation}%(64)
\exp[-2i\theta(\kappa)]=\frac{(\kappa-p_-)(\kappa-p_+)}
{(\kappa+p_-)(\kappa+p_+)}.
\end{equation}
Evidently, there are no zeroes in the upper half-plane for $V(x)\geq 0$. 

\subsection{Exponentially decreasing potentials}
If $V(x)\propto\exp(\mp 2s_\pm x)$ as $x\rightarrow\pm\infty$,
($s_\pm>0$ are constant); $\alpha(\kappa)$ and $\beta(\kappa)$ 
are no longer entire functions. The singularities appear,
when $V(x)$ is not small enough to suppress $\exp(\pm 2i\kappa x)$,
so the integral in Eq. (58) is diverging as $x\rightarrow\pm\infty$.
The singularities  nearest to the real axis appear at 
Im $\kappa=-s_\pm$ for $f$ and $\alpha(\kappa)$, and at
Im $\kappa=\pm s_\pm$ for $g$ and $\beta(\kappa)$,

Explicit expressions for  $a(\kappa)$ and $b(\kappa)$, 
revealing  their singularities, can be obtained, assuming that 
\begin{eqnarray}%(65)
V(x)=v_-^2e^{2s_-(x-x_-)},\;\;x<x_-,\nonumber\\
V(x)=v_+^2e^{-2s_+(x-x_+)},\;\;x>x_+,
\end{eqnarray}
where $v_\pm$ are constants.
The solution to  the Schr\"{o}dinger equation for $x_-<x<x_+$ 
is still a linear combination of $z_0$ and $z_1$, 
while for $x<x_-$ and $x>x_+$ it is given by linear combinations 
of the appropriate Bessel functions. Using the matching conditions
for the wave function at $x_\pm$, one gets,
\begin{eqnarray}%(66)
&&a=-\frac{e^{i\kappa(x_+-x_-)}}{2i\kappa}
\Gamma(1+\nu_+)\Gamma(1+\nu_-)
(\sigma_+/2)^{-\nu_+}(\sigma_-/2)^{-\nu_-}
\nonumber\\
&&\left[\zeta_0' J_{\nu_+}(\sigma_+)J_{\nu_-}(\sigma_-)
+iv_+\zeta_0 {J'}_{\nu_+}(\sigma_+)J_{\nu_-}(\sigma_-)\right. \nonumber\\
&&\left.+iv_-\zeta_1'J_{\nu_+}(\sigma_+) {J'}_{\nu_-}(\sigma_-)
-v_+v_-\zeta_1{J'}_{\nu_+}(\sigma_+){J'}_{\nu_-}(\sigma_-)
\right], \nonumber\\
&&b=\frac{e^{i\kappa(x_++x_-)}}{2i\kappa}
\Gamma(1+\nu_+)\Gamma(1-\nu_-)
(\sigma_+/2)^{-\nu_+}(\sigma_-/2)^{\nu_-}
\nonumber\\
&&\left[\zeta_0' J_{\nu_+}(\sigma_+)J_{-\nu_-}(\sigma_-)
+iv_+\zeta_0 {J'}_{\nu_+}(\sigma_+)J_{-\nu_-}(\sigma_-)\right. \nonumber\\
&&\left.+iv_-\zeta_1'J_{\nu_+}(\sigma_+){J'}_{-\nu_-}(\sigma_-)
-v_+v_-\zeta_1{J'}_{\nu_+}(\sigma_+){J'}_{-\nu_-}(\sigma_-)
\right],
\end{eqnarray}
where $\nu_\pm=-i\kappa/s_\pm$ and $\sigma_\pm=iv_\pm/s_\pm$.
The singularities are due to the $\Gamma$ functions, as soon as
$(\sigma/2)^{-\nu}J_{\nu}(\sigma)$ is known\cite{olver} 
to be an entire function in both $\sigma$ and $\nu$,
while $\zeta$ and $\zeta'$ are analytical functions of $\kappa^2$, 
by the Poincar\'{e} theorem. 
Thus,both $a(\kappa)$ and $b(\kappa)$ have infinite series of equidistant 
poles on the imaginary axis: at $\kappa=-ins_\pm$ for $a(\kappa)$ 
(the poles are double if $s_-=s_+$), and at $\kappa=\mp ins_\pm$ for
$b(\kappa)$ (where $n$ is any positive integer). 

The minimal distance of the singularities from the real $\kappa$-axis,
that was derived from the integral equations, is non-zero for every potential
with an asymptotic exponential decline. The results of Eq. (66)
are less general. If the assumption of Eq. (65) is relaxed, the poles 
can move off the imaginary axis, as one can see in Eq. (72) below.

\subsection{Singularities of the $S$ matrix}
The singularities of the $T$ and $S$ matrices 
are of physical importance, when the time dependent process is
considered. They are given by zeroes of $a(\kappa)$ 
as well as by the poles of $b(\kappa)$, which do not coincide with those
of $a(\kappa)$. 

As was proven in section 4.2,
for positive and integrable potentials, $a(\kappa)$
is analytical in the upper half $\kappa$-plane, which is a result of
causality; $\beta(\kappa)$, and thus $b(\kappa)$,
may have singularities at finite distances above and below the 
real $\kappa$-axis.

The pattern of
singularities of $a(\kappa)$ and $b(\kappa)$ in the complex 
$\kappa$-plane is determined by the asymptotic decline of the potential. 
For potentials, having an asymptotic 
decline faster than exponential, (e.g. finite range potentials,
and the Gaussian barrier), no singularities appear for finite $\kappa$.  
For potentials with exponential asymptotics, 
the distance from the real $\kappa$-axis to the nearest 
singularities is determined by the slope of the potential at $\pm\infty$.

It is important that $a(\kappa)$ has no zeroes
in the upper half plane. As soon as the function is analytical,
the number of its zeroes is given by the integral
   \begin{equation}%(67)
N=\frac{1}{2\pi i}\oint_C\frac{da}{a},
   \end{equation}
where the contour $C$ encloses the upper half of the complex plane.
For positive and integrable potentials $\alpha(\kappa)$ is limited in the 
upper half plane, so $a(\kappa)\rightarrow 1$ as $\kappa\rightarrow\infty$, 
and the integral is zero.

The location of zeroes of $a(\kappa)$ 
in the lower half of the complex $\kappa$-plane, depends on the 
specific barrier considered. For finite-range potentials 
the asymptotic distribution of zeroes is given by Eq. (64).
Other examples are considered in the appendix.

\section{Conclusion}
We have considered the one-dimensional problem, assuming that
the potential is non-negative everywhere. The barrier transmission
and reflection amplitudes are described in terms of two analytical
functions $\alpha(\kappa)$ and $\beta(\kappa)$, Eqs. (29-31).
Both the functions are entire if the potential vanishes
outside a finite interval on the $x$-axis. For potentials decreasing
exponentially, singularities appear at  finite distances to
the real axis, corresponding to the decrease rates at $\pm\infty$.
For $\alpha(\kappa)$, all the singularities are in the lower half-plane,
while for $\beta(\kappa)$, the front slope controls the singularities
in the upper half-plane, and the back slope controls those in the lower
half. Poles of the $S$-matrix are given by zeroes of
$a(\kappa)\equiv 1-\alpha/2i\kappa$. It is proven that for any
non-negative potential they are all in the lower half-plane.

The causality in the transmission and reflection processes manifests 
itself in the analytical properties of the transition amplitudes. 
These properties have been employed for the space-time description of the 
tunneling through potential barrier in the Wigner phase-space 
representation\cite{marsegev96}.

{\em Acknowledgements}.
We are grateful to S. A. Gurvitz and L. P. Pitaevsky for their interest
to this work. The paper was completed when one of the authors (M. M.)
was visiting Max-Planck-Institut at Munich.
It is a pleasure to thank Prof. J. Wess for the kind hospitality.
The support to the research from G. I. F.
and the Technion V. P. R. Fund is gratefully acknowledged.

%%%%%%%%Appendix
\section*{Appendix: Examples}
A number of examples may be found in standard textbooks, e.g.
\cite{flugge}.

i) {\em Square barrier:} $V(x)=V_0$ for $|x|<x_0$,
$V(x)=0$ for $|x|>x_0$.
    \begin{eqnarray}%(68)
a(\kappa)=\frac{1}{4\kappa p}\left[
(\kappa+p)^2e^{2i(\kappa-p)x_0}-
(\kappa-p)^2e^{2i(\kappa+p)x_0}\right] ,\nonumber\\
\beta=V_0\frac{\sin 2x_0 p}{p},
    \end{eqnarray}
where $p=\sqrt{\kappa^2-V_0}$. One can see that these functions
depend, actually, on $p^2$, so there is no cut in the
$\kappa$-plane. Zeroes of $a$ are given by (complex) solutions of
the equation
  \begin{equation}%(69)
e^{-4i p x_0}=\left(\frac{\kappa-p}{\kappa+ p}\right)^2.
   \end{equation}
This equation has no roots for Im$\kappa>0$. If $V_0$ is small enough,
there are two roots on the imaginary $\kappa$-axis. Other roots appear in
pairs, and their asymptotical position is given by
   \begin{equation}%(70)
{\rm Re }p =\pm\frac{\pi}{2x_0}n,\;\;\;
{\rm Im }p=-\frac{1}{x_0}\log\frac{2{\rm Re }p}{V_0},
   \end{equation}
where $n\gg 1$ is an integer.

ii) {\em Exponential barrier:} $V(x)=V_0\exp(-|x/x_0|)$,
    \begin{eqnarray}%(71)
a=-\frac{[\Gamma(1+\nu)]^2}{x_0\kappa}\left(\frac{z}{2i}\right)^{1-4\nu}
J'_\nu(z)J_\nu(z),\nonumber\\
b=\frac{2\pi\sqrt{V_0}}{\sinh2\pi\kappa}
[J'_\nu(z)J_{-\nu}(z)+J_\nu(z)J'_{-\nu}(z)],
  \end{eqnarray}
where $J_\nu(z)$ is the Bessel function, and
$\nu=-2i\kappa$, $z=2i x_0 \sqrt{V_0}$.

iii) {\em The P\"{o}schl -- Teller barrier:} $V(x)=V_0/\cosh^2(x/x_0)$.
  \begin{equation}%(72)
a=i\frac{\Gamma^2(1-i\kappa x_0)}
{\kappa x_0\Gamma(\frac{1}{2}+i\sigma-i\kappa x_0)
\Gamma(\frac{1}{2}-i\sigma-i\kappa x_0)},
\;\;\;b=-i\frac{\cosh\pi\sigma}{\sinh\pi\kappa x_0},
   \end{equation}
where $\sigma=\sqrt{V_0x_0^2-\frac{1}{4}}$. The function $a(\kappa)$
has zeroes at $\kappa x_0=-i(n+\frac{1}{2})\pm\sigma$,
and (double) poles at $\kappa x_0=-i(n+1)$, while $\beta$ has (simple)
poles at $\kappa x_0=\pm in$, $n=1,2,\cdots$. (Note that $\sigma$
is imaginary for $2x_0 < {V_0}^{-1/2}$.)

iv) {\em Narrow barrier:} $V(x)=v_0\delta(x)$.
This is the limit one gets
as $V_0\rightarrow\infty$, $x_0\rightarrow 0$, $2x_0V_0=v_0$, from
the three preceding cases. Now both the entire functions are just constant,
  \begin{equation}
\alpha=\beta=v_0,
   \end{equation}
and $a(\kappa)$ has one zero at $\kappa=-i v_0/2$.

v) {\em A double barrier:} $V(x)=V_1(x-d_1)+V_2(x-d_2)$. 
The case of two non-overlapping barriers
is described by Eq. (35); $\alpha$ and $\beta$ remain entire functions.
As is well known, new zeroes
may appear in $a(\kappa)$ close to the real axis, corresponding
to metastable states of the particle trapped between the barriers.
A special case is that of a symmetrical double barrier,
where $a_1=a_2\equiv\cosh\rho e^{-i\delta}$ and 
$b_1=-\bar{b}_2\equiv\sinh\rho e^{i\gamma}$. Now
   \begin{eqnarray}%(74)
a=\cosh^2\rho e^{-2i\delta}+\sinh^2\rho e^{2i\kappa d},\nonumber\\
b=i\sinh 2\rho\cos(\kappa d+\delta+\gamma),
   \end{eqnarray}
where $d$ is the distance between the barrier centers. It is easy to see  
that the reflection may vanish at certain resonance values of the energy,
independently of the reflection from a single barrier.

%%%%%References

\end{document}